\documentclass[11pt]{article}\textwidth 6.5in\textheight 9in
\usepackage{amssymb}\usepackage[colorlinks]{hyperref}\usepackage{color}
	\usepackage{stmaryrd}\usepackage{mathrsfs}
	\usepackage{graphicx}
	\usepackage{amsmath}
\usepackage{caption}

\begin{document}
\setlength{\captionmargin}{27pt}
\newcommand\hreff[1]{\href {http://#1} {\small http://#1}}
\newcommand\trm[1]{{\bf\em #1}} \newcommand\emm[1]{{\ensuremath{#1}}}
\newcommand\prf{\paragraph{Proof.}}\newcommand\qed{\hfill\emm\blacksquare}

\setcounter{tocdepth}{3} 

\newtheorem{thr}{Theorem} 
\newtheorem{lmm}{Lemma}
\newtheorem{cor}{Corollary}
\newtheorem{con}{Conjecture} 
\newtheorem{prp}{Proposition}

\newtheorem{blk}{Block}
\newtheorem{dff}{Definition}
\newtheorem{asm}{Assumption}
\newtheorem{rmk}{Remark}
\newtheorem{clm}{Claim}
\newtheorem{example}{Example}

\newcommand{\ab}{a\!b}
\newcommand{\yx}{y\!x}
\newcommand{\yux}{y\!\underline{x}}

\newcommand\floor[1]{{\lfloor#1\rfloor}}\newcommand\ceil[1]{{\lceil#1\rceil}}

\newcommand{\lea}{<^+}
\newcommand{\gea}{>^+}
\newcommand{\eqa}{=^+}

\newcommand{\lel}{<^{\log}}
\newcommand{\gel}{>^{\log}}
\newcommand{\eql}{=^{\log}}

\newcommand{\lem}{\stackrel{\ast}{<}}
\newcommand{\gem}{\stackrel{\ast}{>}}
\newcommand{\eqm}{\stackrel{\ast}{=}}

\newcommand\edf{{\,\stackrel{\mbox{\tiny def}}=\,}}
\newcommand\edl{{\,\stackrel{\mbox{\tiny def}}\leq\,}}
\newcommand\then{\Rightarrow}

\newcommand\km{{\mathbf {km}}}\renewcommand\t{{\mathbf {t}}}
\newcommand\KM{{\mathbf {KM}}}\newcommand\m{{\mathbf {m}}}
\newcommand\md{{\mathbf {m}_{\mathbf{d}}}}\newcommand\mT{{\mathbf {m}_{\mathbf{T}}}}
\newcommand\K{{\mathbf K}} \newcommand\I{{\mathbf I}}

\newcommand\II{\hat{\mathbf I}}
\newcommand\Kd{{\mathbf{Kd}}} \newcommand\KT{{\mathbf{KT}}} 
\renewcommand\d{{\mathbf d}} 
\newcommand\D{{\mathbf D}}

\newcommand\w{{\mathbf w}}
\newcommand\Ks{\Lambda} \newcommand\q{{\mathbf q}}
\newcommand\E{{\mathbf E}} \newcommand\St{{\mathbf S}}
\newcommand\M{{\mathbf M}}\newcommand\Q{{\mathbf Q}}
\newcommand\ch{{\mathcal H}} \renewcommand\l{\tau}
\newcommand\tb{{\mathbf t}} \renewcommand\L{{\mathbf L}}
\newcommand\bb{{\mathbf {bb}}}\newcommand\Km{{\mathbf {Km}}}
\renewcommand\q{{\mathbf q}}\newcommand\J{{\mathbf J}}
\newcommand\z{\mathbf{z}}

\newcommand\B{\mathbf{bb}}\newcommand\f{\mathbf{f}}
\newcommand\hd{\mathbf{0'}} \newcommand\T{{\mathbf T}}
\newcommand\R{\mathbb{R}}\renewcommand\Q{\mathbb{Q}}
\newcommand\N{\mathbb{N}}\newcommand\BT{\{0,1\}}
\newcommand\FS{\BT^*}\newcommand\IS{\BT^\infty}
\newcommand\FIS{\BT^{*\infty}}\newcommand\C{\mathcal{L}}
\renewcommand\S{\mathcal{C}}\newcommand\ST{\mathcal{S}}
\newcommand\UM{\nu_0}\newcommand\EN{\mathcal{W}}

\newcommand{\supp}{\mathrm{Supp}}

\newcommand\lenum{\lbrack\!\lbrack}
\newcommand\renum{\rbrack\!\rbrack}

\renewcommand\qed{\hfill\emm\square}

\title{\vspace*{-3pc} How to Compress the Solution}

\author {Samuel Epstein\footnote{JP Theory Group. samepst@jptheorygroup.org}}

\maketitle
\begin{abstract}
Using derandomization, we provide an upper bound on the compression size of solutions to the graph coloring problem. In general, if solutions to a combinatorial problem exist with high probability and the probability is simple, then there exists a simple solution to the problem.  Otherwise the problem instance has high mutual information with the halting problem.
\end{abstract}
\section{Introduction}
In mathematics, the probabilistic method is a constructive method of proving the existence of a certain type of mathematical object. This method, pioneered by Paul Erd\"{o}s, involves choosing objects from a certain class randomly, and showing objects of a certain type occur with non-zero probability. Thus objects of a certain type are guaranteed to exist. For more information about the probabilistic method, we refer readers to \cite{AlonSp04}. Recent results have shown that there is a strong connection between probabilistic method and the compression sizes of encodings of mathematical objects, i.e. their Kolmogorov complexity, $\K$:
\begin{quote}
\textit{If the probabilistic method can be used to prove the existence of an object, then bounds on its Kolmogorov complexity can be proven as well.}
\end{quote}
If there is a simple probability such that objects of a certain mathematical type occur with large probability, then there exists an object of that type that is simple, i.e. has low Kolmogorov complexity. More formally, if object $x$ has $P$-probability of at least $p$ of randomally occuring, then
$$\K(x) \lel \K(P) - \log p + \epsilon.$$
The $\epsilon$ term is the amount of information that an encoding of the entire mathematical construct has with the halting sequence, which can obviously considered to be a negligible amount, except for exotic cases.

This inequality occurs through the application of the EL Theorem \cite{Levin16,Epstein19}. Producing bounds of the Kolmogorov complexity of an object through probabilistic means is called derandomization. In \cite{EpsteinDerandom22}, derandomization was applied to 22 examples, including a number of games. In \cite{EpsteinResource22}, derandomization was used to show the tradeoff in the capacity of classical channels and codebook complexity. In addition, time-resource bounded derandomization was introduced.

I'd recommend derandomization as an area of research for masters students or researchers who are interested in moving into algorithmic information theory. This is because the majority of the technical effort resides in the domain to which derandomization is applied.

In this paper, we show a canonical derandomization example, that of graph vertex coloring. The proof requires an invocation of the EL theorem, an invocation of a conservation theorem, and some straightforward probabilistic arguments.

\section{Results}

The function $\K(x|y)$ is the conditional prefix Kolmogorov complexity. Algorithmic probability is $\m(x)=\{2^{-\|p\|}:U(p)=x\}$, where $U$ is the universal Turing machine. The function $\m$ is universal, in that for any computable probability $P$ over $\FS$, $O(1)\m(x)> 2^{-\K(P)}P(x)$.
Thus for set $D\subseteq\FS$, computable probability $P$, $O(1)\m(D)>2^{-\K(P)}P(D)$. Information between $a\in\FS$ and the halting sequence, $\ch\in\IS$, is $\I(a;\ch)=\K(a)-\K(a|\ch)$, where $\ch\in\IS$ is the halting sequence. ${\lea} f$ is ${<}f{+}O(1)$ and ${\lel} f$ is ${<}f {+} O(\log(f{+}1))$. The following lemma is an information non-growth law. There are many such laws, this one is over asymmetric information with the halting sequence.

\begin{lmm}[\cite{EpsteinDerandom22}]$ $\\
	\label{lmm:consH}
\noindent	For partial computable $f:\N\rightarrow\N$, for all $a\in\N$, $\I(f(a);\mathcal{H})\lea\I(a;\mathcal{H})+\K(f)$. 
\end{lmm}
The following result is the EL Theorem \cite{Levin16,Epstein19}. It was originally formulated as a statement about learning. However since that time, there has been several unexpected applications. In this paper, the EL Theorem is used for derandomization
\begin{thr}[EL Theorem]
\label{thr:el}
For finite $D\subset\FS$, $-\log\max_{x\in D}\m(x)\lel -\log\m(D)+\I(D;\ch)$.
\end{thr}

For graph $G=(V,E)$, with undirected edges, a $k$-coloring is a function $f:V\rightarrow \{1,\dots,k\}$ such that if $(v,u)\in E$, then $f(v)\neq f(u)$. One example of graph coloring is cell phone towers which each need to operate at a certain frequency. Cell phone towers which are two close together cannot have the same frequency due to interference. Thus towers can be represented by vertices, and edges represent interference, with a graph coloring representing an assignment of frequencies to towers. However, what happens if one wants to transmit a description of a proper coloring? The following theorem, a product of derandomization, shows that if more colors are used, the length of an optimal encoding of a proper coloring is smaller.
\begin{thr}
For graph $G=(V,E)$, $|V|=n$ with max degree $d$, there is a $k$ coloring $C$ with $2d\leq k$, and $\K(C) \lel \K(n,k)+2nd/k+ \I((G,k);\ch)$. 
\end{thr}
\begin{prf}
Assume that each vertex is randomly given a color in $\{1,\dots,k\}$. The probability that each vertex does not have a conflict with the other vertices is $\geq\frac{k-d}{k}$. Thus, the probability that the uniform random color assignments is a proper coloring is $\geq\left(\frac{k-d}{k}\right)^n$. Let the finite set $D\subset\FS$ represent all encodings of proper coloring. Thus there is a simple total computable function $f$ that on input $G$ and $k$ can output $D$, with $\K(f)=O(1)$. Let $P$ the uniform probability over all color assignments to the vertices of $G$ (even ones that are not a proper coloring). Thus, bearing in mind that $2d\leq k$,
$$
-\log P(D)< -n\log \left(1-\frac{d}{k}\right) \leq \frac{2nd}{k}.
$$
Thus, due to the definition of the universal semi-measure $\m$,
$$-\log\m(D)\lea \K(P)+P(D)$$
Thus by Theorem \ref{thr:el} and Lemma \ref{lmm:consH}, there is a coloring $C\in D$ with 
\begin{align*}
\K(C)&\lel -\log \m(D) +\I(D;\ch)\\
&\lel \K(P)-\log P(D)+\I(D;\ch)\\
&\lel \K(n,k)+2nd/k + \I((G,k);\ch).
\end{align*}\qed
\end{prf}

\section{Discussion}

Future work involves finding instances of the probabilistic method and applying derandomization to them. In particular, the Lov\'{a}sz Local Lemma, \cite{ErdosLo75}, has been particularly compatible with derandomization. We present the first proved consequence of LLL and show how it is compatible with three versions of derandomization, one that involves Kolmogorov complexity, one that involves resource boundedd Kolmogorov complexity, and one involving games. 

A \textit{hypergraph} is a pair $J=(V,E)$ of vertices $V$ and edges $E\subseteq \mathcal{P}(V)$. Thus each edge can connect $\geq 2$ vertices. A hypergraph is $k$-\textit{regular} of the size $|e|=k$ for all edges $e\in E$. A 2-regular hypergraph is just a simple graph. A valid $C$-\textit{coloring} of a hypergraph $(V,E)$ is a mapping $f:V\rightarrow \{1,\dots,C\}$ where every edge $e\in E$ is not \textit{monochromatic} $|\{f(v):v\in e\}|>1$. The following classic result is proven using LLL.\\

\noindent\textbf{Theorem. (Probabilistic Method)}\textit{
	Let $G=(V,E)$ be a $k$-regular hypergraph. If for each edge $f$, there are at most $2^{k-1}/e-1$ edges $h\in E$ such that $h\cap f\neq\emptyset$, then there exists a valid 2-coloring of $G$.}\\

We can now use derandomization, to produce bounds on the Kolmogorov complexity of the simpliest such 2-coloring of $G$.\\

\noindent\textbf{Theorem A. (Derandomization)}\textit{
	Let $J=(V,E)$ be a $k$-regular hypergraph with $|E|=m$. If, for each edge $f$, there are at most $2^{k-1}/e-1$ edges $h\in E$ such that $h\cap f\neq\emptyset$, then there exists a valid 2-coloring $x$ of $J$ with
	$$\K(x)\lel \K(n)+4me/2^{k}+\I(J;\ch).$$}

We can now use resource derandomization, from \cite{EpsteinResource22} , to achieve bounds for the smallest time-bounded Kolmogorov complexity $\K^t(x) = \min\{p:U(p)=x\textrm{ in $t(\|x\|)$ steps}\}$ of a 2-coloring of $J$.\\

\noindent\textbf{Assumption.} \textit{\textbf{Crypto} is the assumption that there exists a language in $\mathbf{DTIME}(2^{O(n)})$ that does not have size $2^{o(n)}$ circuits with $\Sigma_2^p$ gates.}\\

\noindent\textbf{Theorem B. (Resource Bounded Derandomization)}\textit{ Assume \textbf{Crypto}. Let $J_n=(V,E)$ be a $k(n)$-regular hypergraph where $|V|=n$ and $|E|=m(n)$, uniformly polynomial time computable in $n$. Furthermore, for each edge $f$ in $J_n$  there are at most $2^{k(n)-1}/e-1$ edges $h\in E$ such that $h\cap f\neq\emptyset$.  Then there is a polynomial $p$, and  a valid 2-coloring $x$ of $J_n$ with
	$$\K^p(x)< 4m(n)e/2^{k(n)}+O(\log n).$$ }

We define the following game involving hypergraphs that is from \cite{EpsteinGame23}. The player has access to a list of vertices and the goal of the player is to produce a valid 2-coloring of the hypergraph. We assume that for each edge $f$ of the graph, there are at most $2^{k-1}/e-1$ edges $h$ such that $f\cap h\neq\emptyset$.

The game proceeds as follows. For the first round, environment gives the number of vertices to the player. The player has $n$ vertices, each with starting color $1$. At each subsequent turn, the environment sends to the player the edges which are monochromatic. The player can change the color of up to $k$ vertices and sends these changes to the environment. The game ends when the player has a valid 2-coloring of the graph.\\

\noindent\textbf{Theorem C. (Game Derandomization)} \textit{For $k\geq 6$, there exists a player $\mathbf{p}$ that can beat the environment $\mathbf{q}$ in $(1+\epsilon)n/k$ turns, with Kolmogorov complextiy $\K(\mathbf{p})\lel \I(\mathbf{q};\ch)-\log \epsilon$, where $\epsilon\in(0,1)$.}

\end{document}